\documentclass[twocolumn, prx, superscriptaddress,notitlepage]{revtex4-2}
\usepackage{graphicx}
\usepackage{dcolumn}
\usepackage{bm}
\usepackage[usenames,dvipsnames]{color}
\usepackage{multirow}
\usepackage{gensymb}
\usepackage[normalem]{ulem}
\usepackage{CJK}
\usepackage{comment}
\usepackage[colorlinks, linkcolor=blue,anchorcolor=blue,citecolor=blue,urlcolor=blue]{hyperref}
\usepackage{amssymb}
\usepackage{pifont}
\usepackage{physics}
\usepackage{natbib}
\usepackage{xcolor}

\usepackage{color,soul}

\begin{document}
\begin{CJK*}{UTF8}{}

\title{Ion sensors with crown ether-functionalized nanodiamonds}

\author{Changhao Li}
\thanks{The authors contributed equally to this work.}
\thanks{Current address: Global Technology Applied Research, JPMorgan Chase, New York, NY 10017 USA}
\affiliation{
   Research Laboratory of Electronics, Massachusetts Institute of Technology, Cambridge, MA 02139, USA}
\affiliation{
   Department of Nuclear Science and Engineering, Massachusetts Institute of Technology, Cambridge, MA 02139, USA}

\author{Shao-Xiong Lennon Luo}
\thanks{The authors contributed equally to this work.}
\affiliation{Department of Chemistry and Institute for Soldier Nanotechnologies, Massachusetts Institute of Technology, Cambridge, MA 02139, USA}

\author{Daniel M. Kim}
\affiliation{
   Research Laboratory of Electronics, Massachusetts Institute of Technology, Cambridge, MA 02139, USA}

\author{Guoqing Wang}
\affiliation{
   Research Laboratory of Electronics, Massachusetts Institute of Technology, Cambridge, MA 02139, USA}
\affiliation{
   Department of Nuclear Science and Engineering, Massachusetts Institute of Technology, Cambridge, MA 02139, USA}

\author{Paola Cappellaro}\email[]{pcappell@mit.edu}
\affiliation{
   Research Laboratory of Electronics, Massachusetts Institute of Technology, Cambridge, MA 02139, USA}
\affiliation{
   Department of Nuclear Science and Engineering, Massachusetts Institute of Technology, Cambridge, MA 02139, USA}
\affiliation{Department of Physics, Massachusetts Institute of Technology, Cambridge, MA 02139, USA}

\begin{abstract}
Alkali metal ions such as sodium and potassium cations play fundamental roles in biology. Developing highly sensitive and selective methods to both detect and quantify these ions is of considerable importance 
 for  medical diagnostics and bioimaging.  
Fluorescent nanoparticles have emerged as powerful tools for nanoscale imaging, but their optical properties need to be supplemented with specificity to particular chemical and biological signals in order to provide further information about biological processes.  
Nitrogen-vacancy (NV) centers in diamond are particularly attractive as  fluorescence markers, thanks to their optical stability, biocompatibility and further ability to serve as highly sensitive quantum sensors of temperature, magnetic and electric fields in ambient conditions.
In this work, by covalently grafting crown ether structures on the surface of nanodiamonds (NDs), we build sensors that are capable of detecting specific alkali ions such as sodium cations. We will show that the presence of these metal ions  modifies the charge state of NV centers inside the ND, which can then be  read out by measuring their photoluminescence spectrum. Our work paves the way for designing selective biosensors based on NV centers in diamond.
\end{abstract}

\maketitle
\end{CJK*}

\section{Introduction}
Alkali ions such as sodium and potassium play an essential role in biological systems and their concentration is tightly regulated and  vary in different bodily fluids. For example, sodium and potassium ions are responsible for maintaining fluid and electrolyte balance: the  typical Na$^+$ concentration is 135-150 mM in human blood and only less than 30 mM  in intracellular fluid,   while these proportions are inverted for K$^+$ -- below 5 mM in blood and about 150 mM in intracellular fluid~\cite{SodiumReview2017,MetalIonDetectionReview2020}.  
Fluctuations in their concentrations are usually problematic and can 
lead to various physiological disorders and diseases, including cardiovascular disease and hypertension~\cite{SodiumBBA2010,Viera2015PotassiumDH}. Measurement of their concentrations would thus be of great interest both in understanding the functions of the ions for studying cellular physiology and in clinical examinations. 

Clinical laboratories typically use ion-selective electrodes~\cite{ISEreview2010} or flame photometry~\cite{FlamePhotoReview1945} to perform measurements, but these techniques require a sample volume as large as several mL. To overcome this issue while achieving a high spatial and temporal resolution, small molecule-based fluorescence sensors would be favorable. 
Developing small molecular or nanoparticle probes for sodium or potassium ions has seen substantial advances
in recent years~\cite{MolecularSensorReview2000,SodiumReview2017,MetalIonDetectionReview2020,MetalpHsensingReview2015}. Unlike instrumentation methods in clinical labs, these probes could not only provide a good spatio-temporal resolution but also potentially cross the cell membrane and be used for intracellular measurements. 

\begin{figure*}[thbp]
 \includegraphics[width=1.0\textwidth]{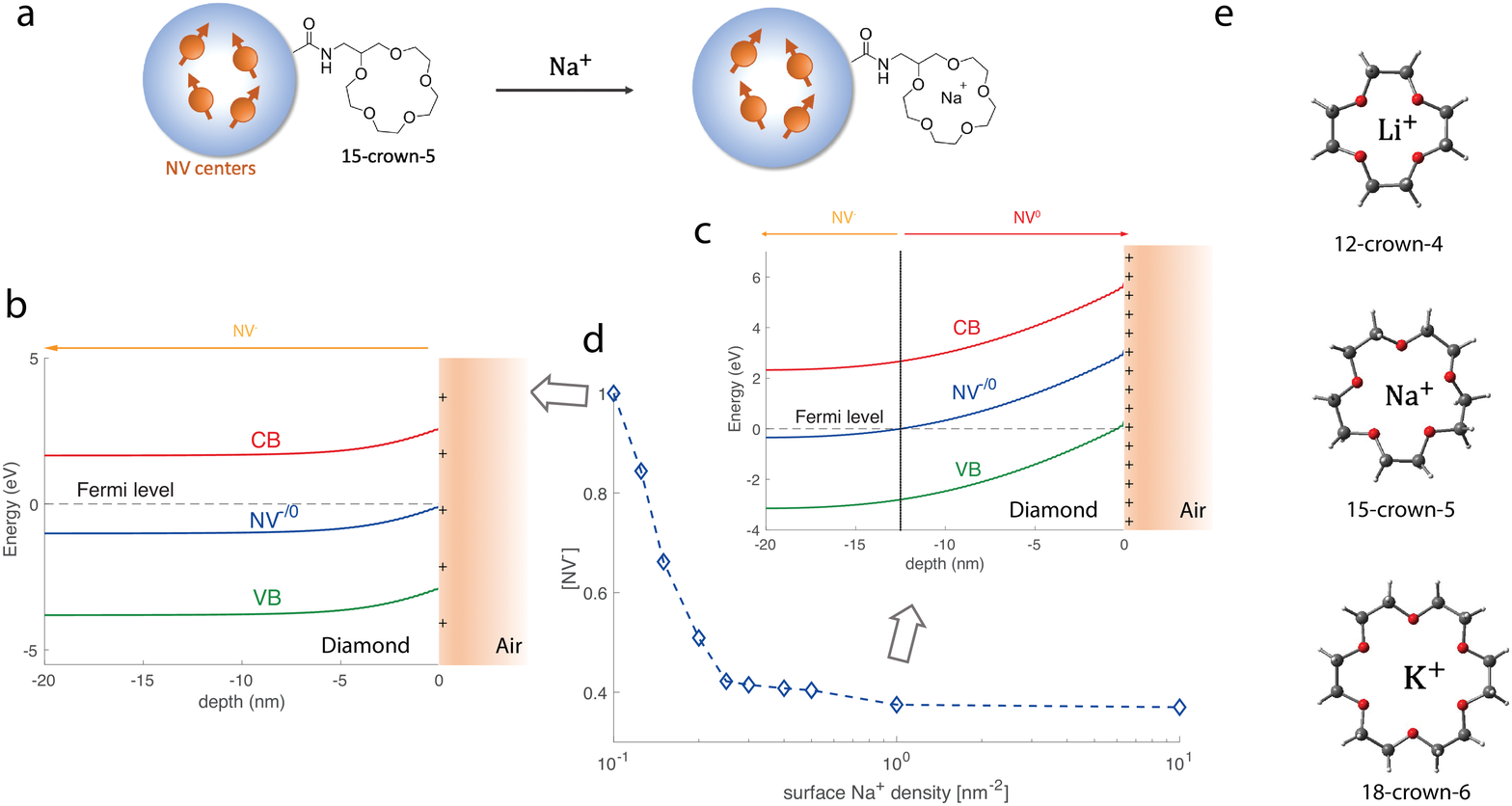}
\caption{\label{fig:fig1} 
\textbf{a.} Schematics of the ion sensor. NV center-containing NDs  are firstly coated with crown ether. Upon their introduction, sodium cations will be trapped in the 15-crown-5 compounds.  \textbf{b.} Energy band schematic of diamond-air interface. Positive charges on the diamond surface lead to upward bending of the conduction band (CB) and valence band (VB) as well as the NV$^{-/0}$ transition level. The dashed line indicates the Fermi level.  NV$^0$ is the dominant NV charge state when the NV$^{-/0}$ transition level is above the Fermi level, while NV$^-$ dominates in the opposite case. The surface density of Na$^+$ is taken to be 0.1 nm$^{-2}$ here. \textbf{c.} Energy band structure when  the surface density of Na$^+$ is  1 nm$^{-2}$ \textbf{d.} Simulated fraction of negatively charged NV states versus surface Na$^+$ density. \textbf{e.} Generalization into other type of crown ethers, which can selectively bind certain cations and form complexes.
}
\end{figure*}

One of the key challenges for designing such ion sensors is to distinguish the target from other common metal ions. As an example, a sodium sensor should be able to tell the difference between sodium (Na$^+$) and other high-concentration ions in biological system, such as potassium (K$^+$), magnesium (Mg$^{2+}$) and calcium (Ca$^{2+}$) ions. In this context, the crown ether family has been widely used in designing alkali metal sensors thanks to its capability of binding alkali ions selectively~\cite{CEreview2004}. Crown ethers are cyclic chemical compounds that consist of a ring containing several ether groups, which can bound specific cations inside the ring. For example, 18-crown-6 is known to have a high affinity for potassium cation whereas 15-crown-5 for sodium cation and 12-crown-4 for lithium cation. Using crown ether together with a fluorescent moiety, highly sensitive metal ion sensor can then be  built~\cite{CEreview2004,CEreview2019}. 
However, the fluorescent dye molecules used by most of the existing metal ion sensors can suffer from photobleaching and fast efflux from cells after loading. Nanoparticles, on the other hand, might provide stable fluorescent signal and last long in cells~\cite{NanotubeSensor2008,DendrimerNanoprobeACSNano2012}. In particular, nanodiamonds (NDs) are bio-compatible and can have stable fluorescence due to inner spin defects, the most-studied one of which is the nitrogen-vacancy (NV) center. The NV center is a point defect in diamond lattice  consisting of a substitutional nitrogen atom with an adjacent vacancy. The negatively-charged NV$^-$ defect has a triplet ground state energy with a zero-field splitting of (2$\pi$)2.87 GHz and can show quantum effects even at room temperature. In recent years, the NV system has been widely used in various quantum applications such as quantum sensors~\cite{QuantumSensingRMP,NVsensingReview2014,NVbiosensingReview2021,LiNL2021,LiNL2019,DNAsensorACS2021} and quantum information processors~\cite{NVQuantumComputingReview2013,NVQuantumComputingReview2021}. In particular, the NV defect is known to be highly sensitive to certain external signals such as magnetic fields and surrounding charge environments. Depending on the local charge distributions and external electrical manipulations, the NV state could have dynamical transitions between the widely-studied negative NV$^-$ state and a neutral state NV$^0$~\cite{WeberPNAS2010}. The charge state of NV centers can then serve as a meter to yield the information of its environment~\cite{DNAsensorACS2021,ChargeDynamicsNL2018,AFM2012,VoltageSensingNPhoton2022,GrotzNC2012}.

In this work, we design and demonstrate an alkali ion sensor based on NV centers in crown-ether-functionalized nanodiamonds. As a proof-of-principle example, we use 15-crown-5 to detect sodium ions. The formation of 15-crown-5-Na$^+$ complexes on nanodiamond surface will sufficiently modify the charge environment of the NV centers inside, and thus alter the charge states of NV defects. We thus measure the  photoluminescence (PL)  emission spectrum of NV defects in nanodiamonds under continuous green laser illumination, from which we can extract the fraction of each NV charge state.  We find that the NV's charge state is influenced both by the surface charge profile and by the power of illumination laser. 
To demonstrate the sensor's capability in detecting the ions presence we exploit the emergence of a distinctive  laser power dependence of the NV$^-$ fraction's upon introduction of sodium cations in the sample, that is clearly different from the pure ND dependence.
We further demonstrate the specificity of the NV-based sensor by  several control experiments. 
While we uses 15-crown-5 in our proof-of-principle experiments, we remark that a different type of crown ether structure can be adapted to probe other alkali ions, including lithium, potassium and cesium ions.

\section{Results}
\subsection{Principles of the sensor}
\begin{figure*}[htbp]
\centering \includegraphics[width=1.0\textwidth]{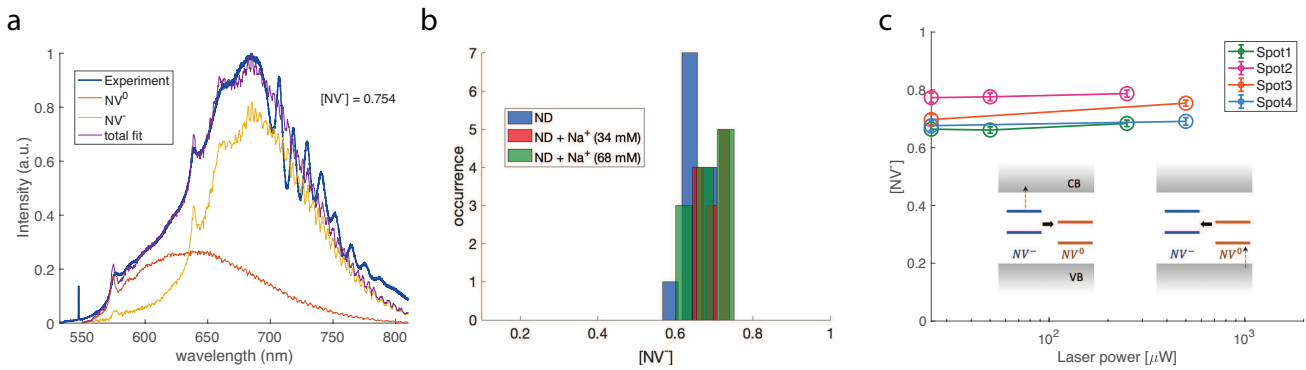}
\caption{\label{fig:fig2} \textbf{a.} Typical  PL spectrum for unfunctionalized NDs (carboxyl-terminated) acquired under 532 nm laser illumination. We perform linear fitting of the spectrum to extract the fraction of NV$^-$ charge state.  The peak-normalized NV$^{-(0)}$ spectrum is from reference~\cite{DirkPNAS2016}. The Raman peak at around 547 nm is attributed to the silicon wafer over which we deposit our samples. Oscillations of PL intensity after 700 nm are due to the etaloning effect on the CCD camera.  \textbf{b.} Distribution of NV$^-$ fraction for carboxyl-terminated NDs in the absence and presence of sodium ions. The laser power is fixed to be 0.5 mW. Sodium cations are added via mixing NaCl solutions with samples of interest hereafter. \textbf{c.} Typical laser power dependence of NV$^-$ fraction of carboxyl-terminated NDs for four different spots. The errorbars are the fitting errors (5 percent). Inset shows the recombination and ionization processes of the two NV charge states.
}
\end{figure*}

We start by introducing the basic detection mechanism of the sensor. As shown in Fig.~\ref{fig:fig1}(a), a fluorescent ND with initial carboxyl group terminations is firstly coated with crown ether via covalent bond. This yields a neutral surface layer in contrast to the negative layer due to the carboxyl groups. Due to the chelate effect and macrocyclic effect, depending on its cavity size, crown ether exhibits strong affinities for specific alkali ions~\cite{CEreview2004,CEreview2019}. 
Upon introducing these ions, stable complexes will form, which  in turn lead to positive charges on the ND surface. For example, the 15-crown-5 structure we studied in this work can strongly bond Na$^+$ and form complexes, while its interaction and affinity with K$^+$ are weak.  While we focus on 15-crown-5 and sodium cation in this work, the scheme is general and can be extended to probe other alkali or alkaline earth metal ions using different types of crown ethers. 
For example, in Fig.~\ref{fig:fig1}(e) we show the case of detecting Li$^+$ and K$^+$.

The accumulated charges on the ND surface can effectively result in bending of the valence band and conduction band in the diamond lattice~\cite{GrotzNC2012,HaufPRB2011,DNAsensorACS2021}. Similar as hydrogen-terminated diamond~\cite{SurfaceCondPRL2000,HaufPRB2011}, as shown in Fig.~\ref{fig:fig1}(b-c), a positive charge layer on the surface can lead to upward bending of both bands and the bending of the NV$^{-/0}$ level, which indicates the energy at which the NV defect loses or takes up one electron. In other words, this transition level indeed corresponds to the  transition from the neutrally to the negatively charged NV states. When its relative position with respect to Fermi level is lower (higher), the defect tends to absorb (lose) one electron.
Due to the positive charge layer on the surface, when close to the surface, this transition level is shifted above the Fermi level (dashed line in Fig.~\ref{fig:fig1}(b-c)) and then the NV defect is ionized from NV$^-$ to NV$^0$. This effect is particularly important for shallow NVs or NVs in ND, while at larger depths the band bending effect is weak and the Fermi level is above the NV$^{-/0}$ level and the dominant charge state would be NV$^-$. We note that the NV$^0$ can further lose an electron and becomes a non-fluorescent state NV$^+$~\cite{WeberPNAS2010}, which is inaccessible here. 

To quantitatively characterize the above model, we perform numerical simulations to study the relative ratio of different NV states (see Methods for detailed information). The NV$^{-/0}$ level corresponds to the ground state of NV$^-$. It's predicted to be in the band gap and it's 2.8 eV above the valence band maximum~\cite{WeberPNAS2010}.  In Fig.~\ref{fig:fig1}(c) we show the simulated band bending effect due to sodium cations on the surface of ND with a diameter of 40 nm when the surface ion density is 1 nm$^{-2}$. At a depth of 12.5 nm, we observe that the dominant species of NV defects switches from neutrally to negatively charged states. We note that close to the surface, the valence band is shifted above the Fermi level, indicating p-type surface conductivity due to the formation of a two-dimensional hole gas.
With Fermi-Dirac distribution, one can then extract the fraction of NV$^-$ as a function of surface cation  density (here we use sodium ion as an example). As expected, in Fig.~\ref{fig:fig1}(d) we observe that a larger density of cations on the ND surface will serve as electron acceptor, thus leading to a smaller NV$^-$ fraction in the diamond lattice.

\begin{figure*}[htbp]
\centering \includegraphics[width=.8\textwidth]{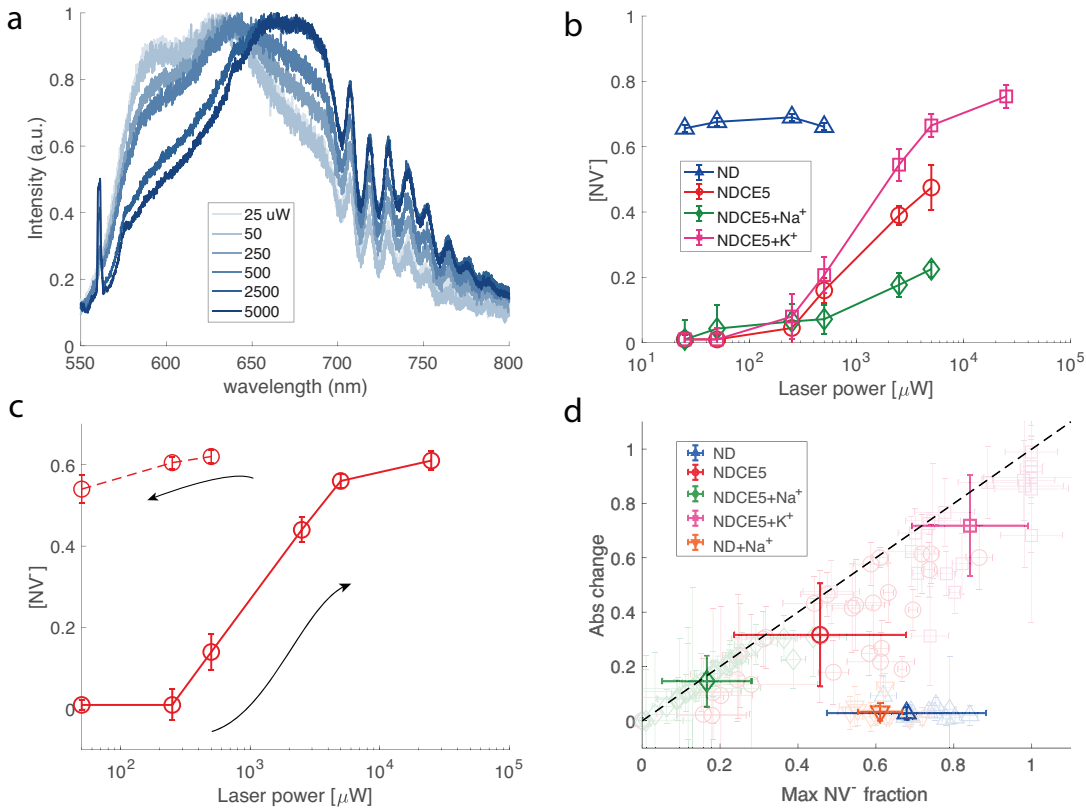}
\caption{\label{fig:fig3} \textbf{a.} PL spectrum of NDCE5 under different illumination powers. \textbf{b.} Typical laser power dependence of NV$^-$ fraction for various samples. Measurements are performed in order of increasing  power. \textbf{c.} Memory curve of NDCE5 sample. The solid curve were first measured in power ascending order and then we went back to low powers (dashed line, in power descending order). \textbf{d.} Comparison of different samples using the coordinates of maximal reachable NV$^-$ fraction and absolute change of NV$^-$ fraction when one varies illumination power. The lighter data points show the results for individual spots, while the opaque data is the average results of different samples. The dashed line is when the change of NV$^-$ fraction equals the maximal, i.e., when starting from [NV$^-$]=0 at low power.
}
\end{figure*}

\subsection{Experiments}

To demonstrate the proposed mechanism for ion sensing, we coat NDs with 15-crown-5 via EDC coupling (see Methods for details) and use the sensor to detect sodium cations.
In order to measure the change of NV$^-$ fraction induced upon the surface charge, we measure the spectrum of nanodiamond ensembles using a Raman spectroscopy (see Methods for details).  NV$^-$ and  NV$^0$ have different PL spectra~\cite{DirkPNAS2016,ScottPRApp2019}, in particular, the zero-phonon-line (ZPL) of NV$^-$ is at 637 nm while it is 575 nm for NV$^0$. To approximately determine the NV$^-$ fraction, we fit the acquired spectrum of our samples $I(\lambda)$  to a linear combination of spectrum of NV$^-$ and NV$^0$:
\begin{equation}
    I(\lambda) = a \{[NV^-]  I_- + (1-[NV^-])  I_0 \},
\end{equation}
where $I_{-(0)}$ denotes the peak-normalized spectrum for a NV$^{-(0)}$ and $a$ is a normalization factor. 
In Fig.~\ref{fig:fig2}(a) we give an example of a typical spectrum for carboxyl-terminated nanodiamonds measured under continuous 532 nm laser illumination. The fitting here yields that the fraction of NV$^-$ is [NV$^-$] = 0.754, which is comparable to the value observed in bulk diamond~\cite{Aslam_2013,ChargeDynamicsNL2018}.

As a control test, we first demonstrate that the charge state of NV centers in the carboxyl-terminated nanodiamonds (labelled as ND in the plots) is not sensitive to sodium cations. First, the ion itself shows no fluorescence under 532 nm laser. Then, as plotted in Fig.~\ref{fig:fig2}(b), the distribution of [NV$^-$] shows little dependence on the concentration of sodium ions. The laser power here is fixed to be 0.5 mW. We further present typical illumination power dependence curves of [NV$^-$] for these unfunctionalized NDs in the absence of external ions (Fig.~\ref{fig:fig2}(c)).
The green excitation dynamically modulates the NV charge state between neutral and negative. 
Due to NV's successive absorption of two photons with wavelength less than 637 nm (corresponds to 1.946 eV), an extra electron of NV$^-$ can be ejected into the conduction band. The first photon pumps the NV$^-$ from ground state to excited state while the second photon takes the electron to the conduction band. In the opposite case, a neutrally charged NV defect can take up one electron from the diamond valence band band via absorption of a photon with energy larger than 2.156 eV (wavelength 575 nm).
Under 532~nm illumination here, both the electron-NV recombination and the NV$^-$ ionization processes can happen, yielding an equilibrium NV$^-$ fraction around 0.7. 
Again, this is similar as NV centers in bulk diamond~\cite{Aslam_2013,ChargeDynamicsNL2018}. To further study the power dependence of the recombination or ionization process separately, another laser with longer wavelength (for example, 632 nm) is required to induce one-directional charge conversion process while suppressing the reverse one.

We then switch to the crown-ether-functionalized nanodiamond sample and measure its fluorescence spectrum under the same 532 nm laser illumination. The measurement is performed from low laser power to high power chronologically. Fig.~\ref{fig:fig3}(a) clearly shows  that now the emission spectrum for 15-crown-5 coated nanodiamond (labelled as NDCE5 hereafter) is power-dependent. When one increases the laser power, the spectrum shifts rightward, corresponding to an increasing NV$^-$ fraction. 
We note that overall the fraction of negative charge state for NDCE5 is smaller than unfunctionalized ND even when the laser power is high. This can be attributed to the fact that NDCE5 is supposed to have neutral charge on the surface, while the carboxyl-terminated NDs tend to have negatively charged surface. 

For a comparison, we then show the the power-dependence of  [NV$^-$] for different samples in Fig.~\ref{fig:fig3}(b).  While for unfunctionalized NDs we observe no dependence on illumination power, for all crown-ether-coated samples studied in this work, we see that when the power is low the NV$^-$ fraction is approximately zero and it then keeps increasing as a function of illumination power. A similar phenomenon has been observed in near-surface shallow NV centers~\cite{ChargeDynamicsNL2018} and this is in contrast to NV centers in certain bulk samples~\cite{ScottPRApp2019}, where the NV$^-$-to-NV$^0$ ratio decreases as a function of the laser intensity. The power dependence observed here emerges from the interplay between the ionization and recombination rates and the charge transfer rate from NV$^-$ to neighboring traps~\cite{ChargeDynamicsNL2018,GWChargeTransport}. These trap states are likely originated from neighboring defects such as vacancy complexes. 

While the NDCE5 curve typically shows a saturated fraction of about 0.5,  after adding sodium cations, the NV$^-$ fraction is considerably lower than 0.5 (usually below 0.4) even at high laser power. We attribute this to  the positive charge on ND surface yielding the band bending effect. 
We further show the data for NDCE5 in the presence of potassium ions. In contrast to sodium ions, 15-crown-5 has low affinity for potassium cation and the sample shows a high [NV$^-$] at large laser power. 

We note that we observe a memory effect for the NDCE5 sample studied above. That is, after finishing the spectrum measurement at high laser power, we set the power to low values and re-evaluate the properties of spectrum. Fig.~\ref{fig:fig3}(c) shows that the NV$^-$ fraction remains at a high value (above 0.5) even when the laser power is set as low as $50~\mu$W. The memory effect persists for at least hours. While a detailed study on the charge dynamics and electron transfer process is needed, we suspect that high laser power preferentially transfers an excess electron from local traps to the NV defect~\cite{ChargeDynamicsNL2018}.

Finally, in Fig.~\ref{fig:fig3}(d) we summarize the results for various samples we evaluated. Here we plot the absolute change of [NV$^-$] as a function of the maximal [NV$^-$] the sample can achieve at high illumination power. We note that in the presence of sodium ions, the maximal fraction is considerably lower than other samples, including NDCE5 and unfunctionalized  NDs. Again, this is due to the fact that 15-crown-5 can form complexes with sodium cations, leading to a positive charge layer on the ND surface. This in turn results in the band bending effect and thus in a lower NV$^-$ fraction even when the laser illumination power is high. Surprisingly, we find that in the presence of K$^+$, at high power most of the NV defects can be converted to NV$^-$. Further study is needed to investigate the interaction between 15-crown-5 and potassium cations.

\section{Discussions}

While our results demonstrate that the NV charge ratio can be used to detect the presence of targeted cations, the large inhomogeneities in the ND properties prevented us from obtaining quantitative results on the [NV$^-$] density. 
As our experimental setup lacked the ability to repeatedly  addressing the same ND, we had to perform measurements over many spatial spots to average out the inhomogeneities among NDs and extract significant differences as a function of sample properties. 
The inhomogeneities in fluorescence intensity, initial charge ratio and ionization(deionization) rates
 are induced by various factors, including inefficient surface coating of crown ether, spatial density profile of NV centers and nitrogen atoms, as well as distinct local charge environments. 
A well-calibrated, single ND sensor would avoid this costly repeated measurement process to extract the behavior distribution,  and  greatly improve the sensitivity of the sensor.  NV centers with pre-characterized local charge environments and surface charge densities might be used to reduce the overhead in measuring many spatial spots. 

While we used an all-optical approach to read out the charge state information of NV defects via their PL spectrum measurement, we remark that one might extract the same information from the signal contrast of optical-detected magnetic resonance (ODMR), given the fact that NV$^0$ and NV$^-$ have distinct ground state spin configurations. A larger NV$^-$ fraction will yield a better signal contrast at the resonance frequency (2.87 GHz in the absence of external magnetic fields). In this case, a microwave pulse would be required for the ODMR experiment. With the help of microwave, the surface charge layer might also be probed by monitoring the transverse relaxation time of the NV$^-$ charge state, as it can induce fluctuating electrical field that can interact with the NV centers  shortening its relaxation time~\cite{NVElectricalPRL2015,NVElectricalNPhy2011}.

In conclusion, we designed a sodium ion sensor based on NV centers in crown-ether-functionalized nanodiamonds. The charge state of NV centers shows a strong dependence on the surface charge profile and can be detected by measuring the PL spectrum of NV centers. While we focused on sodium cations here, the sensing mechanism can be extended to detect other ions such as K$^+$ by changing the surface crown ether structures. These NV-based sensors have a stable PL signal and can provide excellent spatial resolutions, thus opening new opportunities for monitoring ion concentrations in biological systems and for cellular physiology.

\section*{Acknowledgement}
This work was supported in part by the U.S. Army Research
Office through Grant W911NF-15-1-0548 and by the National Science Foundation.

D.K. acknowledges the support from the National Institute of General Medical Sciences with award Number T32GM007753. The content is solely the responsibility of the authors and does not necessarily represent the official views of the National Institute of General Medical Sciences or the National Institutes of Health.

\section*{Competing financial interest}
The authors declare no competing financial interests.

\section*{Author contributions}
C.L. and S.-X. L. L. proposed the sensor scheme and designed the experiment. P.C. supervised the project.   S.-X. L. L. conducted the sensor synthesis and characterization. C.L. performed the PL spectrum measurement and data analysis, with partial input from G.W. The numerical simulations were performed by C.L. and D.K. using the nextnano software.  C.L. wrote the paper with the contributions from all authors.  All authors discussed the results.

\appendix
\section{Experiment details}
\subsection{Nanodiamond sample}
The nanodiamonds  (Adamas Nanotechnologies, NDNV40nmHi10ml) in this work had an average size of 35-40 nm and were milled from high pressure high temperature (HPHT) micro-size particles. Before milling, these particles were irradiated with 2-3 MeV electrons followed by annealing at 850$^{\circ}$C for 2 hrs. Substitutional nitrogen content in the starting micro-size material was around 100 ppm, while the concentration of NV defect is found to be 1-2 ppm, corresponding to around 12-14 color centers per 40 nm nanoparticle~\cite{AdamasDocument2017}.

\subsection{Chemical synthesis and characterization}
\begin{figure*}[htbp]
\centering \includegraphics[width=.9\textwidth]{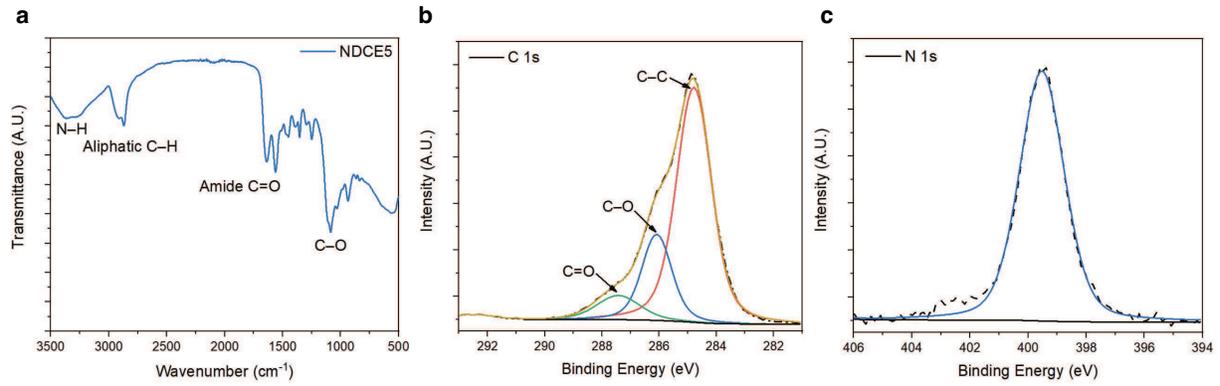}
\caption{\label{fig:sample_XPS_FTIR} Characterization of NDCE5 sample. \textbf{a.} FT-IR spectrum of NDCE5. \textbf{b-c.} XPS C 1s and N 1s spectra of NDCE5.
}
\end{figure*}

Commercial reagents were purchased from Sigma-Aldrich and TCI and used as received unless otherwise noted. Following reported procedures~\cite{YANG2020484,GARG2019172}, nanodiamonds were first treated with a mixture of acids to remove the surface graphite contaminants and metallic impurities, generating carboxyl groups for the subsequent functionalization by EDC coupling. In particular, nanodiamonds were dispersed in a 3:1 mixture of concentrated sulfuric acid and nitric acid and stirred overnight at room temperature. After neutralization with aqueous NaOH (1 M), the resulting nanodiamonds were cleaned using several centrifugation, washing, and redispersion cycles with Milli-Q water. EDC coupling with amino crown ether was performed by adding an excess of 1-(3-dimethylaminopropyl)-3-ethylcarbodiimide hydrochloride and 2-aminomethyl-15-crown-5 to the nanodiamonds dispersion and the reaction mixture was stirred overnight at room temperature, followed by several centrifugation, washing, and redispersion cycles with Milli-Q water. The resulting nanodimonds were characterized by a Bruker Alpha II FTIR spectrometer with a Diamond Crystal ATR (attenuated total reflectance) accessory and a K-alpha+ X-ray Photoelectron Spectrometer system (Thermo Scientific) using a  Al K$\alpha$ radiation source (Fig.~\ref{fig:sample_XPS_FTIR}).

\subsection{Spectrum measurement and fitting}
The photoluminescence (PL) spectrum of the NV centers under 532 nm laser excitation is collected via a confocal Raman microscope (Renishaw inVia microscope with a 1024 $\times$ 256 pixel CCD camera) at room temperature. We put the samples on a silicon wafer to avoid unwanted Raman peaks. 

To approximately extract the fraction of NV$^-$ charge state  we first peak-normalize the measured curves and then perform linear fitting of the spectrum from 560 nm to approximately 750 nm based on the single NV$^{-}$ and NV$^0$ reference spectra~\cite{DirkPNAS2016}.

\section{Simulations}
To qualitatively understand the change of NV charge state in the presence of surface charges,  numerical simulations are performed. 
We first assume rapid thermal equilibrium state after the photoexcitation process of the NV centers~\cite{GrotzNC2012,HaufPRB2011}. The charge state of NV center is then exclusively governed by the relative position of the NV$^{-/0}$ charge transition level with respect to the Fermi level, neglecting the complex ground, excited and metastable states of all defect states.  We first perform numerical simulations of the band bending effect using the nextnano software~\cite{Nextnano2006}, a tool for simulation of electronic semiconductor nanodevices. With realistic parameters, one can use it to calculate the band structure of multi-dimensional devices. 

In this work, we perform three-dimensional simulations on a nanodiamond of spherical shape with a diameter $d=40$ nm. In the simulation, the Poisson equation is discretized on a grid with grid size 0.5 nm using the finite differences method and solved iteratively in a self-consistent manner. The Poisson equation is coupled with the single-band effective mass Schr\"odinger equation via the charge density, as described by the wave functions in the diamond~\cite{GrotzNC2012,HaufPRB2011}. Negative charge (electron) donors in the lattice mainly include the substitutional nitrogen atoms (ionization energy 1.7 eV) and here we take their concentration to be 100 ppm (as expected from the ND used in experiments) and assume uniform distributions as a function of depth below the diamond surface. We take the nitrogen-to-NV conversion ratio to be 1$\%$, corresponding to a total NV defect concentration of 1 ppm. The boundary condition is determined by the surface charges (depth $x=0$) that are contributed by the 15-crown-5-Na$^+$ complexes or 15-crown-5 structures. Considering the strong affinity for sodium cation of 15-crown-5, we assume that the density of surface crown ethers is identical to the density of surface positive ions. 

After getting the relative position of the charge transition level $E_{-/0 }$ with respect to the Fermi level $E_F$, we use the Fermi-Dirac distribution to extract the average fraction of NV$^-$ charge state at depth $x$:
\begin{equation*}
    \langle n_-(x) \rangle = \frac{1}{1+e^{(E_{-/0 } - E_F)/k_B T}} 
\end{equation*}
where $k_B$ is the Boltzmann constant and T = 300 K is the temperature. 
The total fraction of NV$^-$ in the whole nanodiamond lattice with radius $r$ can then be calculated by performing the integral
\begin{equation*}
    [NV^-] = \int_0^r \langle n_-(x) \rangle g(x) dx
\end{equation*}
where $g(x)$ is the normalized NV defect density profile in the diamond lattice, i.e., $\int_0^r g(x) dx = 1$. In our simulations, we assume uniform distributions of NV defects.

\bibliography{PRApp_submit_2301}

\end{document}